# MigrationMiner: An Automated Detection Tool of Third-Party Java Library Migration at the Method Level

Hussein Alrubaye*, Mohamed Wiem Mkaouer*, Ali Ouni†

*Software Engineering Department, Rochester Institute of Technology, NY, USA
†ETS Montréal, University of Quebec, Montréal, QC, Canada
{hat6622,mwmvse}@rit.edu, ali.ouni@etsmtl.ca

*Abstract*—In this paper we introduce, *MigrationMiner*, an automated tool that detects code migrations performed between Java third-party library. Given a list of open source projects, the tool detects potential library migration code changes and collects the specific code fragments in which the developer replaces methods from the retired library with methods from the new library. To support the migration process, MigrationMiner collects the library documentation that is associated with every method involved in the migration. We evaluate our tool on a benchmark of manually validated library migrations. Results show that MigrationMiner achieves an accuracy of 100%. A demo video of Migration-Miner is available at https://youtu.be/sAlR1HNetXc.

## I. Introduction

The tremendous growth of available third-party libraries as being an integral part of modern software ecosystems, engendered new maintenance and evolution challenges. Typical challenges are mainly related to library APIs upgrade and migration as they often get deprecated or outdated. *third-party library migration* [1], [2] is the process of replacing a library with a different one, while preserving the same program behavior. Unlike, upgrading a library from one version to another, the migration requires developers to explore and understand the new library's API, its associated documentation, and its usage scenarios in order to find the right API method(s) to replace every method, belonging to the retired library's API.

Existing studies demonstrated that library migration is still a manual, error-prone, and time-consuming process [3], [4], [5], [6], [7]. Developers often spend a considerable time to check whether the newly adopted features do not introduce any regression in their client code. Indeed, recent studies have shown that developers typically spend up to 42 days to migrate between libraries [8]. Moreover, recent studies have shown that developers are reluctant to migrate their existing libraries, which makes their overall dependencies outdated and even vulnerable [6]. Hence, there is an urgent need to support developers in migrating their third-party libraries.

In this tool paper, we present, MigrationMiner[1], an open source tool that provides the developer with easy-to-use and comprehensive way of extracting, from given list of input projects, existing migrations between two third-party libraries using program analysis based on Abstract Syntax Tree (AST) code representation. In a nutshell, MigrationMiner (*i*) detects, (*ii*) extracts, (*iii*) filters, and (*iv*) collects code changes related to any performed migration. For a given input project, MigrationMiner *detects* any migration undergone between two java libraries and returns the names and versions of both retired and new libraries. Thereafter, MigrationMiner *extracts* the specific code changes, from the client code, and which belong to the migration changes (it should at least have one removed method from the retired library, and one added method from the new library) from all other unrelated code changes within the commits. Next, MigrationMiner *filters* code changes to only keep fragments that contain migration traces *i.e.*, a code fragment, generated by the *diff* utility, which contains the removed and added methods, respectively from the retired and the new library. Finally, MigrationMiner *collects* the library API documentation that is associated with every method in the client code. The output of MigrationMiner, for each detected migration between two libraries, is a set of migration traces, with their code context, and their corresponding documentation.

To the best of our knowledge, there is no available open source tool that can extract migration traces between two different libraries. MigrationMiner is the first initiative to provide an open source tool and a dataset of automatically detected migrations[2]. Developers can use it to outsource from the *wisdom of the crowd*, and extract migration patterns between two given libraries. Thus, developers can use it as a *by-example* approach, to facilitate their migration process. Researchers can use it also to better understand the challenges associated with library migration and get practical insights.

**Tool, documentation and demo video.** Migration-Miner is publicly available as an open source tool[1], with a demo video[3].

## II. Background

This section presents definitions of the main concepts that are used throughout the paper.

---

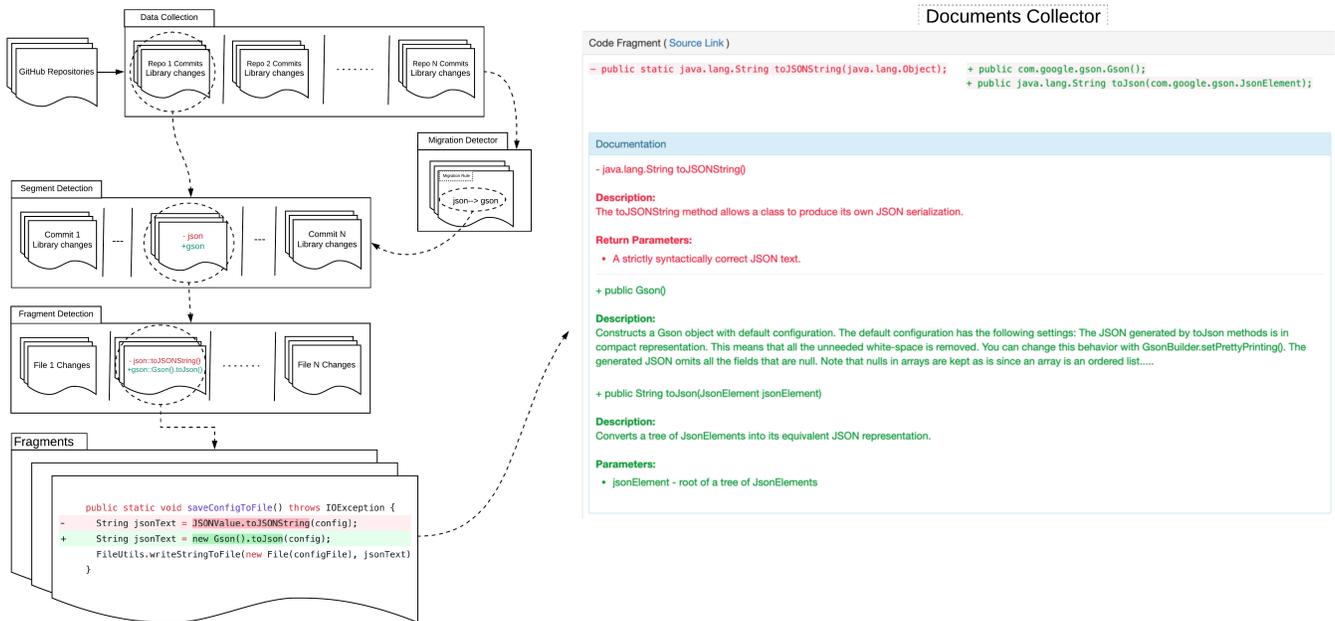

Figure 1: MigrationMiner workflow and Architecture.

**Library Migration.** A library migration occurs when a source library is replaced by a target library. The source library is considered retired if all of its method dependencies are removed from its client code. Note that the source library does not need to be physically removed from the project (*e.g.*, the pom files for Maven projects, or local libraries repository), but it enforces that none of its methods are used in the client implementation.

**Migration Rule.** A migration is denoted by a pair of a source (retired) library and a target (replacing) library, *i.e.*, *source → target*. For example, *json → gson* represents a Migration Rule where the library json[4] is migrated to the new library gson[5].

**Method Mapping.** A Migration Rule is a set of method mappings between the source and the target library. The mapping between methods is the process of *identifying the correspondences* between a least one method from the source library and one or multiple methods belonging to the target library.

**Segment.** It constitutes the migration *period*. It is a sequence of one or multiple code changes (*e.g.*, commits), containing each, one or multiple fragments.

**Fragment.** A block of source code that witnesses at least one mapping. It is generated by contrasting the code before and after the migration to only keep the removed (resp., added) methods linked to the source (resp., target) library.

## III. MIGRATIONMINER

In this section we detail the architecture and typical usage scenario of MigrationMiner as sketched in Figure 1. For each MigrationMiner component, we explain its input/output and workflow.

### A. Data Collection

**Input.** MigrationMiner takes as input a list of open source GitHub projects, as shown in Figure 1. Due to the mining nature of our tool, we allow multiple project links to facilitate the automated search.

**Workflow.** The collection phase takes the list of open-source Java projects. It starts by cloning and checking out all commits for each project. For every commit, MigrationMiner collects its properties including the commit ID, commit date, developer name, and commit description. MigrationMiner keeps track of all changes in the project library configuration file, known as Project Object Model (*pom.xml*). All mined projects data is recorded in a SQL database for faster querying later when identifying segments and fragments. As an illustrative example, Figure 2 shows a commit [6] where *json* was removed from the project while *gson* was added.

**Output.** A list of potential library changes, and their corresponding commits, and projects.

| | |
|---|---|
| - | `<groupId>org.json</groupId>` |
| - | `<artifactId>json</artifactId>` |
| - | `<version>20080701</version>` |
| + | `<groupId>com.google.code.gson</groupId>` |
| + | `<artifactId>gson</artifactId>` |
| + | `<version>2.3.1</version>` |

Figure 2: Migration from *json* to *gson*.

---

[4]https://mvnrepository.com/artifact/org.json/json/20140107

[5]https://mvnrepository.com/artifact/com.google.code.gson/gson/2.8.0

[6]https://github.com/vmi/selenese-runner-java/commit/641ab94e7d014cdf4fd6a83554dcff57130143d3

### B. Migration Detector

**Input.** List of library changes, and their corresponding commits, and projects.

**Workflow.** Since developers may add and remove multiple libraries at the same time, there is no clear cut way to figure out the pairs of removed/added libraries. Therefore, the Cartesian Product (CP) is performed between the set of added removed libraries, in each parsed *pom.xml* files, to extract all the possible combinations between removed/added libraries. Figure 3.A demonstrates the CP process in the form of a graph. Every node in the graph represents a library while the edge represents its potential mapping to another library. The edge is weighted by the number of times a migration is found, while parsing all commits, across all projects. For instance, the edge between *json* to *gson* has a weight of 12 because this migration has been identified 12 times during the data collection process. Since the CP generates every possible combination of rules, its result contains a large number of false positives. Thus, a two-step filtering process is performed:

1) In the first step, as shown in Figure 3.B, the weights are normalized by the highest outgoing weight per node, then the only mappings kept are those with a normalized weight that is higher than a user-defined filtering threshold value $t_{rel} \in [0, 1]$. The value of $t_{rel}$ controls the selection strictness. For example, when the filter $t_{rel} = 1$, the *json* to *gson*, *easymock* to *Mockito*, and *testng* to *Junit*, migration are selected. MigrationMiner has $t_{rel} = 1$ by default, to guarantee a strict selection of rules.

2) The second filtering step is ensured by the *Fragment Detection* component, where only rules with actual migration traces at the method level are kept, *i.e.*, for a rule *source → target*, it is only kept if and only if there exists at least one or many method(s) from *source* that has/have been replaced with one or many methods from *target*. The functionality of this component is detained in Section III-D.

**Output.** Migration Rules with the highest weights.

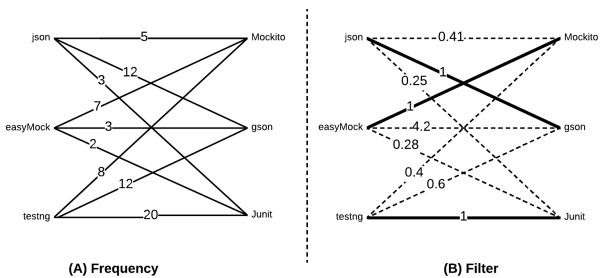

**(A) Frequency**  **(B) Filter**

Figure 3: Library Migration Detector.

### C. Segment Detection

**Input.** Migration Rules with the highest weights. Besides, list of library changes, and their corresponding commits.

**Workflow.** The purpose of the segments detection, *i.e.*, migration periods, is to locate, for each Migration Rule, its time periods in all projects. As defined in the background section, a segment could be composed of one or many commits involved in the migration process. As shown in Figure 1, the segment detection phase starts with checking whether both libraries exist in the list of added/removed project libraries. Using static program analysis, MigrationMiner locates the end of the segment by scanning all commits in which all project source files are no longer dependent on the retired library.

Note that a migration does not require the physical removal of the library to be retired from the project, as the retired library may still loaded in project through its *pom.xml* file; however none of the library's methods are used in the client code. Once a segment *end* is located, MigrationMiner keeps scanning previous commits in a backward fashion, looking for the *start* commit which contains the beginning of the migration, *i.e.*, the first code change related to the replacement of any retired library method. After locating all segments for a given migration, it is important to keep track of source and target libraries versions for each segment to avoid backward incompatibility in case of an API change between two versions of the same library.

**Output.** Migration Rules with the highest weights, and their corresponding segments.

### D. Fragment Detection

**Input.** Migration Rules with the highest weights, and their corresponding segments.

**Workflow.** Fragment detection generates source code fragments related to the library migration changes as shown in Figure 1. It clones the project source files that are changed in the commits belonging to the identified time segments. We apply the Git's *Unified Diff Utility* command between the changed files to generate fragments, if any. A fragment is a continuous set of lines that have been changed along with contextual unchanged lines. Only fragments containing removed (resp., added) methods from the source (resp., target) library are considered valid. For example[7], in Figure 1, for a given Migration Rule *json → gson*, we identify that one of the fragments was converting `object` to `string`. We only keep `toJSONString()` as a removed method and `Gson()`, `toJson(Object)` as an added method. Other code changes (*i.e.*, `String jsonText`) that do not belong to the migration, will be removed. As previously explained, all rules with no found fragment(s), will be automatically discarded.

**Output.** Filtered Migration Rules, and their corresponding fragments.

---

[7]line 180 in RuntimeConfig.java , https://github.com/groupon/Selenium-Grid-Extras/commit/4d9bada8aeab5b09e7a27926fc9ecab8bb5a1b51

### E. Documentation Collector

**Input.** Filtered Migration Rules, and their corresponding fragments.

**Flow.** As shown in Figure 1, for a given fragment, the documentation collector collects the API documentation for both source method(s) and target method(s). Based on its corresponding Migration Rule, it automatically downloads the library documentation as a *jar* file for all library releases involved in migrations. Our approach relies on the libraries documentation on Maven Central Repository[8]. The largest online Java library ecosystem hosting over 3,605,525 unique libraries[9], as of April 2019. The documentation collector then converts the API documentation from a *jar* file to multiple *HTML* source files using the *doclet API*[10]. It parses all of the *HTML* files and collects the documentation related to class descriptions, method descriptions, parameter descriptions, return descriptions, package names, and class names. Finally, the documentation collector identifies the documentation associated with every method involved in any of the Migration Rules. For example, Figure 4 shows *Gsons'* API documentation[11] for the method `String toJson(JsonElement)`.

Figure 4: API documentation for *toJson(JsonElement)*

**Output.** Filtered Migration Rules, their fragments, and the documentation for each method in any fragment.

### IV. Case Study

To evaluate the correctness of our detection process, we challenge our tool using an existing dataset provided by Teyton et al. [2]. This dataset contains 4 Migration Rules and their corresponding method mappings, detected across 16 projects, from which only 7 projects were using Maven, and so compatible with our tool. To challenge the ability of MigrationMiner to identify all the Migration Rules and their related mappings, we consider these 7 projects (that contain 3 migrations). Then, we compare our findings with the results of their manual detection to calculate the precision and recall. As shown in Table I, MigrationMiner was able to detect all the Migration Rules, and all their corresponding fragments, achieving precision and recall of 100%. More interestingly, we have identified three additional Migration Rules, namely $lucene-core \rightarrow compass$, $jmf \rightarrow gstreamer-java$, $jersey-client \rightarrow wink-client$, along with their fragments. We manually inspected and validated all the detected fragments in the client code. Thus, MigrationMiner achieved a precision of 100%. Since we did not manually investigate whether there are more unrevealed fragments, calculating the recall is not applicable for this case study.

Table I: Accuracy of Migration Miner.

| Migration Rule | New? | Precision | Recall |
|---|---|---|---|
| $commons-lang \rightarrow guava$ | No | 100% | 100% |
| $commons-io \rightarrow guava$ | No | 100% | 100% |
| $commons-lang3 \rightarrow guava$ | No | 100% | 100% |
| $lucene-core \rightarrow compass$ | Yes | 100% | N/A |
| $jmf \rightarrow gstreamer-java$ | Yes | 100% | N/A |
| $jersey-client \rightarrow wink-client$ | Yes | 100% | N/A |

### V. Conclusion and Future Work

We presented MigrationMiner, an open source tool to detect migrations between third-party Java libraries. The evaluation of Migration Miner has shown its effectiveness in detecting manually validated migrations. MigrationMiner has been already used to detect and study the 9 various migration, detected in 57,447 projects, and this work has been published in the 27th IEEE/ACM International Conference on Program Comprehension [8]. As future work, we plan to extend MigrationMiner, to provide an interactive tool for the recommendation of library migration at the method level.